# Radiopathomics: Integration of radiographic and histologic characteristics for prognostication in glioblastoma

Saima Rathore[1,2], Muhammad A. Iftikhar[3], Metin N. Gurcan[4], Zissimos Mourelatos[5]

[1]Center for Biomedical Image Computing and Analytics; [2]Department of Radiology, University of Pennsylvania, Philadelphia, PA, USA; [3]Comsats University Islamabad, Lahore Campus, Lahore, Pakistan; [4]Wake Forest School of Medicine, Bowman Gray Center for Medical Education, Winston-Salem, NC; [5]Department of Pathology and Laboratory Medicine, University of Pennsylvania, Philadelphia, PA, USA.

**Abstract**
Both radiographic (Rad) imaging, such as multi-parametric magnetic resonance imaging, and digital pathology (Path) images captured from tissue samples are currently acquired as standard clinical practice for glioblastoma tumors. Both these data streams have been separately used for diagnosis and treatment planning, despite the fact that they provide complementary information. In this research work, we aimed to assess the potential of both Rad and Path images in combination and comparison. An extensive set of engineered features was extracted from delineated tumor regions in Rad images, comprising T1, T1-Gd, T2, T2-FLAIR, and 100 random patches extracted from Path images. Specifically, the features comprised descriptors of intensity, histogram, and texture, mainly quantified via gray-level-co-occurrence matrix and gray-level-run-length matrices. Features extracted from images of 107 glioblastoma patients, downloaded from The Cancer Imaging Archive, were run through support vector machine for classification using leave-one-out cross-validation mechanism, and through support vector regression for prediction of continuous survival outcome. The Pearson correlation coefficient was estimated to be 0.75, 0.74, and 0.78 for Rad, Path and RadPath data. The area-under the receiver operating characteristic curve was estimated to be 0.74, 0.76 and 0.80 for Rad, Path and RadPath data, when patients were discretized into long- and short-survival groups based on average survival cutoff. Our results support the notion that synergistically using Rad and Path images may lead to better prognosis at the initial presentation of the disease, thereby facilitating the targeted enrollment of patients into clinical trials.

## 1. Introduction

Glioblastoma is a frequent and aggressive malignant adult brain tumor, with poor prognosis and heterogeneous molecular[1] and imaging[2] landscape. The median overall-survival of glioblastoma patients is ~14-16 months after standard-of-care treatment, or possibly a bit longer with some emerging advanced treatments[3], and ~4 months otherwise[4]. Although the currently applicable treatment options, which include surgical resection, radiotherapy, and chemotherapy, have expanded during the last couple of decades, the prognosis still remains poor for this devastating disease. A major obstacle in treating glioblastoma patients is the heterogeneity of their molecular and imaging profiles [5,6], which renders the clinical practice of the same treatment for all inadequate.

Both radiographic (Rad) imaging such as multi-parametric magnetic resonance imaging (MRI), and digital pathology (Path) images captured from tissue samples are routinely acquired as standard clinical practice for glioblastoma patients. Both these data streams provide comprehensive information, but are separately used for diagnosis and treatment planning, despite the fact that they provide complementary information. However, the combination of these imaging sequences under the umbrella of emerging field of radiopathomics may lead to better diagnosis compared to either of the individual imaging sequences alone.

Previously, it has been shown that both Rad and Path images based signatures can independently predict outcome of interest. For example, MRI based signatures of epidermal growth factor receptor variant III *(EGFRvIII)* mutation [7], isocitrate dehydrogenase-1 (IDH1) mutational status[8], and $O^6$-methylguanine–DNA methyltransferase *(MGMT)* methylation status [9], and Path based signatures of *IDH* and *1p/19q* mutation [10,11] are indicative of the fact that both these imaging sequences manifest the biology of the tumor and hence lead to prediction of outcome of interest. Likewise, it has also been shown in the recent literature that MRI[12] and Path[10,11,13] can both lead to assessment of survival of brain tumors. Lately, Shukla et al. presented that the features extracted from MRI offer additional predictive value to the survival assessment acquired by using clinical and genomic biomarkers only in the TCGA glioblastoma patients [14]. Results by another group on the same dataset also confirmed better prognostication when Path images and genomic biomarkers (*IDH*, *1p/19q*) were used together[15]. These data and some other studies on prostate cancer [16,17] support the hypothesis that combined evaluation of Rad and Path images will even further improve prognostication, and will enhance our understanding of the disease. Therefore, objective of this study is to apply supervised learning algorithm on multiple data streams to exploit the complementary information provided by these imaging sequences for improved prognostication. We integrated and analyzed the entirety of Rad and Path imaging data, leveraging advanced pattern analysis methods.

These advanced methods enable the extraction of quantitative imaging features and subsequently unique imaging phenotypes, which are hard to appreciate either via visual analysis of images or by analysis of individual features. In this research study, we aim to use advanced image analysis to demonstrate the ability of integrated radiomics and pathomics data for assessment of survival of glioblastoma patients.

## 2. Materials and Methods

### 2.1 Study Setting and Data Source

The data of 107 glioblastoma patients involved in this study was acquired from The Cancer Image Archive (TCIA) and The Cancer Genomic Atlas (TCGA). The criteria to include patients in the study were availability of: (i) pre-operative MRI data, including T1-weighted (T1), T1 post contrast gadolinium (T1-Gd), T2-weighted (T2), and T2 fluid attenuated inversion recovery (T2-FLAIR) images, (ii) corresponding digital pathology images, and (iii) overall-survival outcome. The overall-survival of these patients was downloaded from clinical records provided by TCIA. The patients were acquired under different scanners and image acquisition methods in different institutions, which allowed testing of the robustness of the method across different institutions.

### 2.2 Image Preprocessing

Considering the totally different nature of imaging sequences involved in the study, two separate image processing pipelines were used to process Rad and Path images (Fig.1).

**Rad (MRI) images:** DICOM import routines available in SPM8 were used to convert the MRI images to stereoscopic T1, T2, T2-FLAIR and T1-Gd volumetric images. The MRI image sequences were smoothed in order to reduce intensity-based noise in image regions having uniform intensity profile. An adaptive non-local means algorithm[18,19], which is an extended version of traditional non-local means algorithm, was used for de-noising. The optimal parameters for de-noising were obtained using a previously proposed method[20]. For correction of bias in MRI scans caused due to magnetic field inhomogeneities, N4 bias correction algorithm was used to correct for intensity non-uniformities caused by the magnetic field of the scanner during image acquisition[21]. Registration and skull-stripping were, respectively, performed by the Linear Image Registration Tool (FLIRT) and Brain Extraction Tool (BET)[22]. The histogram matching was performed on the images by selecting all the images of one patient as a reference template.

**Path images:** All the Path images were manually checked for artifacts, and the images free of all types of artifacts were chosen. The images were then converted to gray-scale using matlab routines before further image processing.

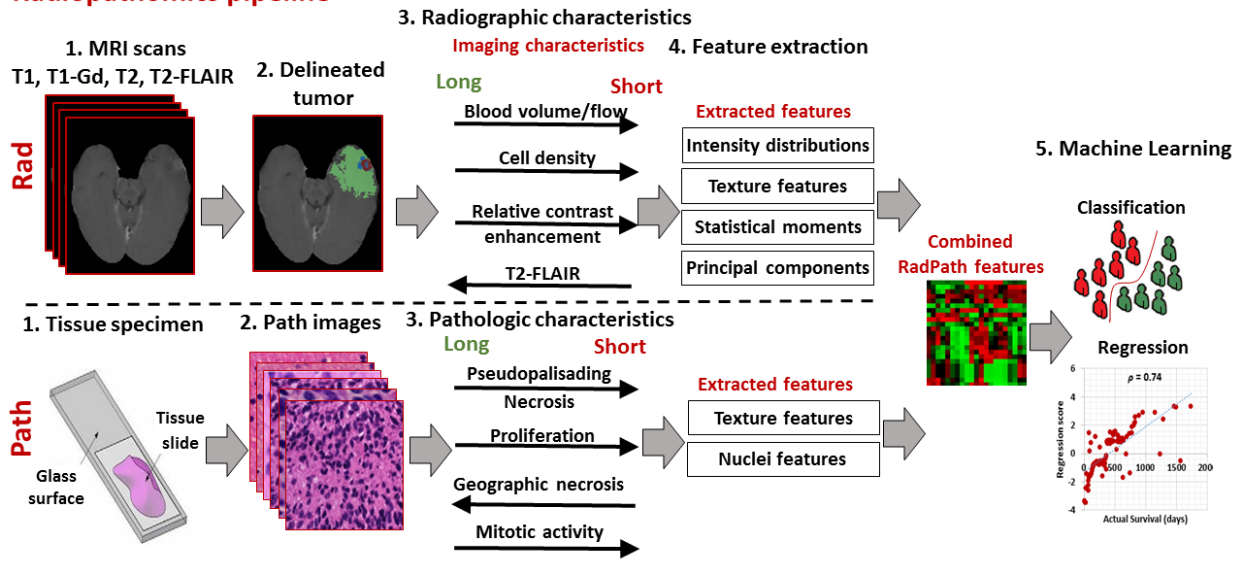

Figure 1: Image processing workflow. 1. Pre-processed MRI images and tissue specimen. 2. Delineation of tumor sub regions in MRI scan, and extraction of random patches from Path image. 3. Radiographic and pathologic characteristics of long and short survivors. 5. Extracted radiomic features from MRI, and radiomic and nuclei features from Path images. 5. Machine learning regression for prediction of overall-survival, and machine learning classification to model survival prediction as a binary classification problem between long- and short-survivors.

## 2.3 Region Annotation

Specific regions were either delineated or extracted from Rad and Path images to derive features for the classification problem.

**Path images:** A set of 100 region-of-interest each comprising 1024x1024 pixels which had viable tissue with vivid histopathologic characteristics and that were free of artifacts were extracted from Path images. The patches were selected in a way to make sure that they cover atleast 50% of the tissue area (Fig. 2a).

**Rad images:** Different tumor sub-regions comprising enhancing tumor (ETumor), nonenhancing tumor (non-ETumor) core, peritumoral edema region or invasion (Edema), and ventricles (Vent) were manually delineated (Fig. 2b).

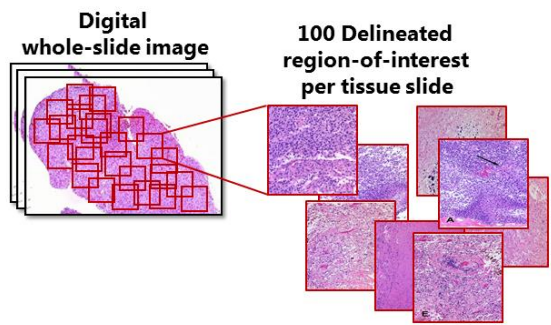

(a)

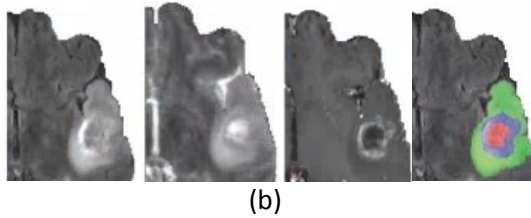
(b)

Figure 2: Region annotation in Path and Rad images: (a) 100 patches of size 1024x1024 pixels were extracted from the whole slide Path image, (b) tumor region delineated into different tumor sub-regions, including ETumor (blue), Non-ETumor (red), and Edema (green).

## 2.4 Feature Extraction

Following the definition of these ROIs, all Rad and Path images were analyzed to extract relevant comprehensive quantitative imaging phenomics features from the corresponding ROIs, in order to create our predictive model.

**Path images:** A reconstruction algorithm based on fast hybrid grayscale was employed to normalize the background region and to get rid of the artifacts introduced by the scanning and tissue cutting process[23]. A simple threshold-based mechanism was employed to extract the foreground nuclei from the normalized background region, followed by watershed segmentation algorithm for proper delineation of overlapped nuclei[24]. Four complementary types of features were extracted from segmented nuclei. The features include size, morphometry, nuclear intensity and gradient statistics, and texture descriptors summarizing the content and distribution of chromatin within nuclei. In addition, a comprehensive set of traditional radiomic texture features such as local binary patterns (LBP)[25], gray-level size-zone matrix (GLSZM)[26,27], neighborhood gray-tone difference matrix (NGTDM)[28], gray-level co-occurrence matrix (GLCM)[29], and gray-level run-length matrix (GLRLM)[26] were also extracted.

**Rad (MRI) images:** Several features were extracted from each delineated tumor sub-region i.e. Edema, Non-ETumor, and ETumor. The features included: volumetric measures of tumor sub-regions; anatomical location of the tumor; distance, in millimeter, of ETumor and Edema from ventricles; first order statistical moments of intensities of each Rad sequence in Edema, Non-ETumor, and ETumor; frequency of intensities of each Rad sequence in five equal-sized distribution bins in ETumor, non-ETumor, and Edema; the summarized signal of the intensity distribution histogram of each Rad sequence using Principal Component Analysis (PCA); age; and gender. A comprehensive set of traditional radiomic texture features, as extracted from Path images, including LBP[25], NGTDM[28], GLSZM[26,27], GLCM[29], GLRLM[26] was also extracted. To obtain these texture features in 3 dimensions, all mpMRI volumes were first quantized to 16 gray levels within the ROI. A neighborhood of 3x3x3 was considered for GLCM. These features were first computed for each of the 13 main directions independently, and then averaged to find their final value. All features were rescaled via z-score normalization before further analysis.

## 2.5 Machine learning and correlation analysis

Support Vector Machines (SVM), a pattern classification approach, was used to construct a classifier, to predict long- and short-survivors. Linear kernel of SVM was employed for this multivariate analysis, and the parameters of the classifier were optimized using cross-validated grid search (5-fold) on the training data. The classifier was trained separately using leave-one-out cross-validation (LOOCV) schema within training and validation sets and each time one of the three types of features was used, i.e., Rad, Path and RadPath. To confirm the robustness, accuracy, and generalizability of the method in a larger cohort, while avoiding optimistically biased estimates of performance, we have also evaluated the classifiers in all 107 patients using a LOOCV schema, and in a split-train-test mechanism by training the classifier on the training cohort (n=54) and testing on the validation cohort (n=53).

In addition to identifying an imaging signature to distinguish between long- and short-survivors, we also tried to find the correlations between the features and the overall-survival. This approach should identify complementary information of the extracted features by their correlations with long- and short-survivors. To achieve this, we used all 107 patients and trained separate support vector regression (SVR) models in a LOOCV configuration for the prediction of survival with continuous values.

## 3. Results

### 3.1 Performance of the proposed radiopathomics regression model

Pearson correlation coefficient ($\rho$) and area under the curve (AUC) were employed for evaluation. A boost in performance [$\rho$=0.79, AUC=0.768] was observed using RadPath images compared to either Rad [$\rho$=0.75, AUC=0.76] or Path [$\rho$=0.75, AUC=0.73] features on combined dataset in a LOOCV setting (top row Fig.3). The regression models based on Rad and Path images were able to yield [$\rho$=0.73, AUC=0.76] and [$\rho$=0.74, AUC=0.78], respectively, in split-train-test configuration. However, the performance improved considerably when the features extracted from both the image types were used together in the regression model. A higher correlation of 0.79 and AUC 0f 0.80 indicates that a significant improvement in survival prediction is achieved when both the image types are used together (bottom row Fig.3). Also, Similarly, the regression models developed using RadPath features on training and validation cohorts were able to yield [$\rho$=0.81, AUC=0.79] and [$\rho$=0.80, AUC=0.79], respectively. The corresponding performance was [$\rho$=0.74, AUC=0.78] and [$\rho$=0.75, AUC=0.73] using Rad, and was [$\rho$=0.74, AUC=0.73] and [$\rho$=0.73, AUC=0.70] using Path images.

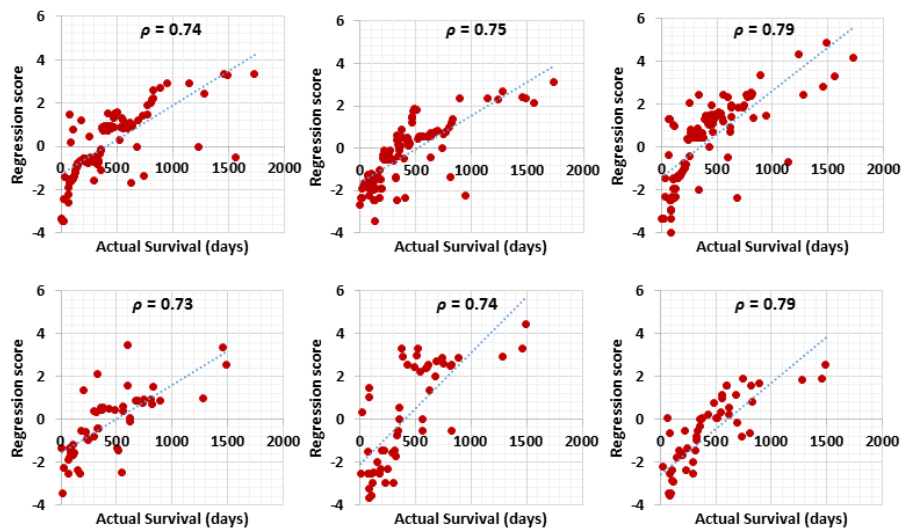

Figure 3: Correlation analysis of the scores of different experiments: Top-row and bottom row, respectively, show classification results for LOOCV on complete dataset and the split-train-test experiment. Left to right show Rad, Path, and RadPath features, respectively.

### 3.2 Performance of the radiopathomics classification model

The accuracy of the model in predicting long and short survivors was 81.82% in the training cohort, 80.77% in the validation cohort, and 82.24% in the combined cohort using Rad features only in a LOOCV setting (results given in Table 1). The cross-validated classification performance of the model was 85.45% in the training cohort, 82.69% in the validation cohort, and 85.05% in the combined cohort using Path features only. The performance further improved with RadPath features yielding an accuracy of 90.91% in the training cohort, 86.54% in the validation cohort, and 90.65% in the combined cohort. Overall, 2-way classification into short and long survivors was 75.00%, 78.85%, and 84.62% using Rad, Path, and RadPath features, respectively, in the split-train-test setting. The corresponding ROC curves are in Fig.4.

Table 1: Quantitative evaluation of the proposed model using Rad, Path and RadPath feature sets in various classification settings: LOOCV in training, validation and combined datasets, and a model applied on replication data after being trained on the training dataset.

|  | Accuracy | Sensitivity | Specificity | AUC, S.E. [95% CI] |
|---|---|---|---|---|
| **Rad** | | | | |
| LOO (Training) | 81.82 | 82.14 | 81.48 | 0.84, 0.06 [0.73-0.95] |
| LOO (Validation) | 80.77 | 80.77 | 80.77 | 0.80, 0.06 [0.68-0.92] |
| LOO (Combined) | 82.24 | 81.48 | 83.02 | 0.83, 0.04 [0.75-0.91] |
| Split-Train-Test | 75.00 | 73.08 | 76.92 | 0.72, 0.07 [0.58-0.86] |
| **Path** | | | | |
| LOO (Training) | 85.45 | 82.14 | 88.89 | 0.84, 0.06 [0.73-0.95] |
| LOO (Validation) | 82.69 | 80.77 | 84.62 | 0.83, 0.06 [0.72-0.95] |
| LOO (Combined) | 85.05 | 85.19 | 84.91 | 0.83, 0.04 [0.75-0.91] |
| Split-Train-Test | 78.85 | 76.92 | 80.77 | 0.74, 0.07 [0.60-0.88] |
| **RadPath** | | | | |
| LOO (Training) | 90.91 | 89.29 | 92.59 | 0.89, 0.05 [0.80-0.98] |
| LOO (Validation) | 86.54 | 84.62 | 88.46 | 0.89, 0.05 [0.80-0.98] |
| LOO (Combined) | 90.65 | 90.74 | 90.56 | 0.88, 0.03 [0.82-0.95] |
| Split-Train-Test | 84.62 | 84.62 | 84.61 | 0.86, 0.05 [0.76-0.97] |

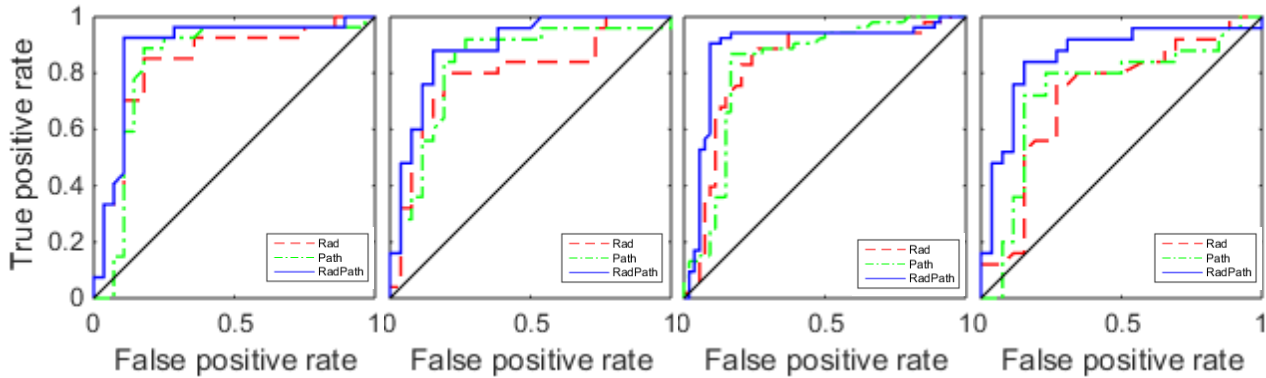

Figure 4: ROC curves of different experiments using Rad, Path, and RadPath features. Left to right are LOO (training), LOO (Validation), LOO (Combined), and split-train-test.

### 3.3 Survival analysis

Kaplan–Meier survival curves for both the groups (long- and short-survivors) based on the predictions of the proposed model are shown in Fig.5. Both cross-validated model on complete data, and a model applied on the validation data after being trained on the training cohort are included. The patients predicted to be long- and short-survivors by the model had higher and lower actual survival, respectively. The hazard ratio between the predicted long- and short-survivors was 2.867 (95% confidence interval (CI) = 1.921-4.278, P < .001), 2.736 (95% CI = 1.831-4.087, P < .001), and 4.33 (95% CI = 2.846-6.587, P < .001) for Rad, Path and RadPath datasets, respectively, in the complete cohort. The corresponding hazard in the split-train-test setting was 2.665 (95% CI = 1.489-4.771, P < .001), 2.525 (95% CI = 1.403-4.544, P < .001) and 5.972 (95% CI = 2.991-11.93, P < .001) for Rad, Path and RadPath datasets, respectively.

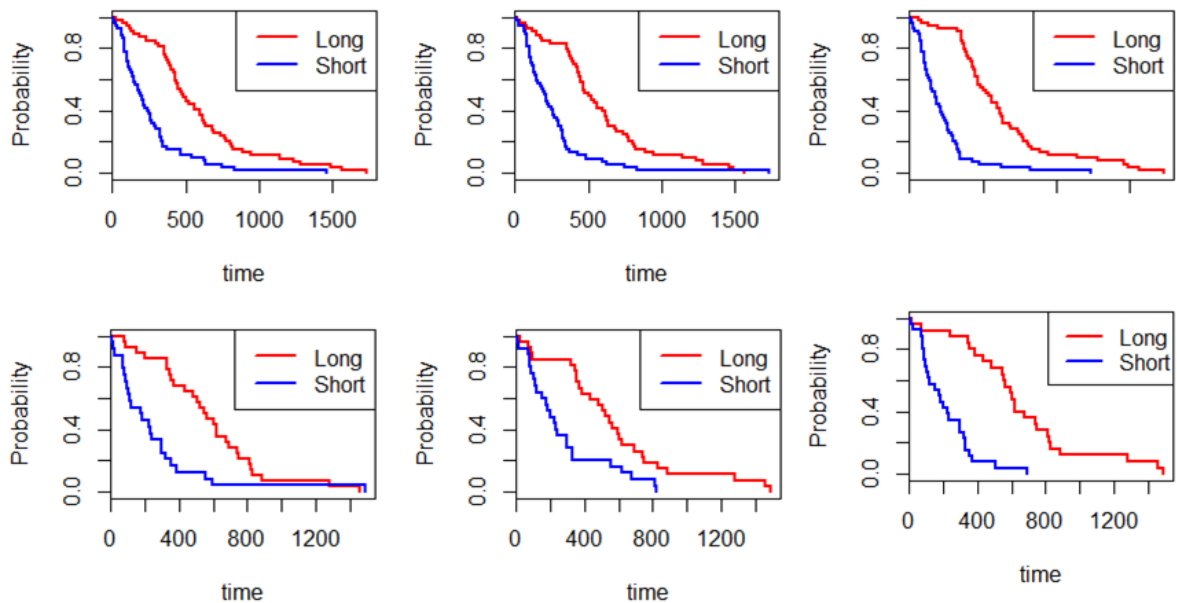

Figure 5: Kaplan–Meier survival curves. X-axis shows the actual in days and y-axis shows the predictions generated by the proposed model. Top-row and bottom row, respectively, show classification results for LOOCV on complete dataset and the split-train-test experiment. Left to right show Rad, Path, and RadPath features, respectively.

## 4. Discussion

There has been substantial interest in the last decade in 'radiogenomics' and 'pathogenomics' methods where the aim has been to correlate radiological and pathological image features with genomic profiles of tumors. Additionally, there is mounting recent interest in developing methods for automatically co-registering *ex-vivo* pathology with *in-vivo* radiographic imaging in order to spatially map disease extent from the pathology onto the radiographic imaging[16,17].

Alterations in gene expression implore changes at the vascular and structural level in phenotype of tissue that sequentially can be detected on the particular imaging modality under observation. For example, morphology of the tumor as seen in tissue slides mimics the aggregate effect of molecular alterations/pathways in tumorous cells. Likewise, radiographic imaging sequences highlight structural and functional characteristics of the tissue, which reflect biological pathways and cellular morphology portraying the tumor cells. Hence, having a panel of computational tools combining histology and radiographic sequences is likely to improve our ability to target the right patients with the right treatments, and to monitor response over the whole course of the disease.

In this article, we employed advanced pattern analysis methods in a cohort of glioblastoma patients that have undergone *in-vivo* radiographic and *ex-vivo* digital pathology imaging. We identified a RadPath based imaging signature of the long- and short-survivor glioblastoma patients, and studied biological correlates of the topmost distinctive radiographic and pathologic features. Most importantly, the proposed imaging signature was derivative of a comprehensive and diverse panel of morphological and physiological characteristics of the tumors extracted from radiographic and pathologic images. The individual features are not sufficient for identifying long- and short-survivors on a single patient basis; but, suitable combination of these features through machine learning accurately identifies long- and short-survivors on an individual patient basis, underscoring the value of using a RadPath model that synergistically uses all the imaging features extracted from Rad and Path images.

The proposed method underscores the clinical workflow of radiologists and pathologists by developing predictors on radiology and pathology images separately, and by providing integrated diagnosis which may lead to interconnected effort between various department for the management of glioblastoma patients. In terms of Path images, digitization of tissue slides facilitates the real-time transmission of information-rich digital pathology images between different facilities for research, diagnostics, and tutoring purposes. This is mainly suitable for acquiring second-opinion on difficult cases and the option to provide remote consultation without physically shipping tissue slides across different facilities. The digitization of tissue slides may also advance clinical workflow by minimizing the requirement of storing glass slides in bio-banks of pathology departments and decreasing the risk of glass tissue slides getting damaged. In addition to substantially aiding the radiologists and pathologists in their reporting workflow and clinical decision making, the computational tools could facilitate the development of imaging based supporting diagnostic assays that could not only allow for improved risk characterization of a disease but also enable the integration of diagnostic data coming from different departments for better disease management.

In terms of Rad images, availability of such biomarkers can contribute to non-invasive measurement of molecular profile on individual-patient basis. The RadPath approach has the added benefit of both the approaches. Contrasting to costly molecular based assays that not only destroy the tissue and assess molecular markers from a tiny fraction of the tumor, thereby underestimating tumor heterogeneity, these pathology and radiographic imaging (RadPath) based companion diagnostic tools could be made available at very reduced price, and could facilitate characterization of the heterogeneity of brain tumors across the entire breadth of the tissue specimen or radiographic appearance of the tumor. These tools can help in patient stratification into appropriate treatments, and identification of patients with relatively highly heterogeneous tumors, who would benefit from more extensive histopathological and molecular analysis through multiple samples, as well as by combination treatments.

Our study is most likely to be immediately translatable to routine clinical settings and to contribute to precision medicine owing to the use of standard imaging modalities being routinely acquired in almost all the institutions. Moreover, the proposed approach has been evaluated on glioblastoma patients, but is generalizable enough to be applicable to other cancer types, especially other brain tumors. Since radiographic and pathology imaging both capture spatial heterogeneity of the tumor across the entire landscape of the tumor, and are used over time (especially radiographic imaging) to evaluate the treatment response in glioblastoma, the proposed RadPath approach for stratification can possibly contribute throughout the course of the treatment of the patients. Furthermore, as the field of medicine is evolving and the trend towards team-based management of diseases is rapidly increasing, improved communication and information exchange between different departments such as radiology and pathology is needed more than ever. The proposed RadPath platform provides a mechanism to generate coherent, correlated, and integrated diagnostic summaries with nominal additional effort from radiologists and pathologists. We believe that a rich set of radiology/pathology features, and the machine learning signatures derived from these will enhance our understanding of glioblastoma and can contribute to precision diagnostics.

This work can be extended along various lines: 1) detailed analysis of the discriminative features of both the imaging sequences that regression model selects in various cross-validation loops, 2) use of advanced deep learning method for jointly using both the imaging sequences, and 3) to perform histology-radiology correlation based on the histology-radiology features most discriminative in determining the survival outcome.

## 5. Conclusion

The current standard of diagnosis for brain tumors includes uncoordinated communication/transfer of diagnosis reports from radiology and pathology departments to the treating physicians. Any disagreement or conflict between these results generally requires added time from the treating physicians to correctly diagnose the disease, or given incorrect diagnosis, may unfavorably lead to wrong or sub-optimal treatment decisions. To overcome these problems, we proposed and developed a system (pipeline), called RadPath, for combining radiology and pathology images for integrated diagnostics. We in fact developed a new automated method that synergistically utilizes radiology and pathology imaging sequences, in combination and comparison, of glioblastoma patients for assessment of survival outcome of these patients. The results of the proposed method on a dataset comprising 107 subjects are quite encouraging and prove the effectiveness of the proposed method not only in terms of combined imaging sequences, but also on individual imaging sequences as well.